\documentclass[letterpaper, 10 pt, conference]{ieeeconf}  % Comment this line out if you need a4paper

\IEEEoverridecommandlockouts                              % This command is only needed if 
\usepackage{amsmath,amssymb,amsfonts}
\usepackage{algorithmic}
\usepackage{graphicx}
\usepackage{textcomp}
\usepackage{xcolor}
\usepackage{adjustbox}
\usepackage{multirow}%
\usepackage{comment}
\usepackage{graphicx}
\usepackage{subcaption}
\usepackage{float}
\usepackage{url}
\usepackage[noadjust]{cite}
\def\BibTeX{{\rm B\kern-.05em{\sc i\kern-.025em b}\kern-.08em
    T\kern-.1667em\lower.7ex\hbox{E}\kern-.125emX}}

\title{\LARGE \bf
Designing Empathetic Companions: Exploring Personality, Emotion, and Trust in Social Robots
}

\author{Alice Nardelli*$^{1}$, Antonio Sgorbissa$^{1}$, Carmine Tommaso Recchiuto$^{1}$
\thanks{* Corresponding author. {\tt\small alice.nardelli@edu.unige.it/}}% <-this % stops a space
\thanks{$^{1}$ DIBRIS Department, RICE laboratory, University of Genoa
        {\tt\small https://rice.dibris.unige.it/}}%
}

\begin{document}

\maketitle
\thispagestyle{empty}
\pagestyle{empty}

%%%%%%%%%%%%%%%%%%%%%%%%%%%%%%%%%%%%%%%%%%%%%%%%%%%%%%%%%%%%%%%%%%%%%%%%%%%%%%%%
\begin{abstract}
How should a companion robot behave? In this research, we present a cognitive architecture based on a tailored personality model to investigate the impact of robotic personalities on the perception of companion robots. Drawing from existing literature, we identified empathy, trust, and enjoyability as key factors in building companionship with social robots. Based on these insights, we implemented a personality-dependent, emotion-aware generator, recognizing the crucial role of robot emotions in shaping these elements. We then conducted a user study involving 84 dyadic conversation sessions with the emotional robot Navel, which exhibited different personalities. Results were derived from a multimodal analysis, including questionnaires, open-ended responses, and behavioral observations. This approach allowed us to validate the developed emotion generator and explore the relationship between the personality traits of Agreeableness, Extraversion, Conscientiousness, and Empathy. Furthermore, we drew robust conclusions on how these traits influence relational trust, capability trust, enjoyability, and sociability.

\end{abstract}


\section{Introduction}
%need of companion robot
A companion robot is designed to facilitate specific activities while establishing emotional connections \cite{ahmed2024human}. Properly designing robot behavior has been shown to be fundamental for building companionship across various applications, e.g., education \cite{zinina2023learning}, %\cite{reich2015learning}, 
healthcare \cite{lu2021effectiveness}, and %\cite{cooper2020ari}, 
entertainment \cite{hoffman2016robotic}. %, loneliness mitigation %\cite{odekerken2020mitigating}, 
%\cite{berridge2023companion}, and shopping \cite{bertacchini2017shopping}. 
Many of these applications rely on \textit{Empathy} and \textit{Trust}, which are recognized as key factors in building companionship, by creating an emotional connection between the human and the robot \cite{ahmed2024human}.

%robotic personality
Designin \textit{robot personality}, on the other hand, has been identified as a successful strategy for enhancing several aspects of Human-Robot Interaction (HRI), including likability, enjoyment, knowledge acquisition, engagement, trust, and warmth %\cite{peeters2006personality}, 
%\cite{robert2018personality}, 
\cite{esterwood2021systematic}, \cite{robert2020review},
\cite{rossi2020role}, \cite{galatolo2023personality}. 
%link between robotic personality and trust and empathy
A few studies have explored the possible effects of 
%\textit{Human personality} and \textit{Trust} \cite{oksanen2020trust}, \cite{rossi2018impact}. On the other side, 
 \textit{robot personality} on \textit{trust}. For example, it has been shown that language style, shaped by personality, can influence trust \cite{lim2022we}, \cite{robb2023explanation}, and that modulating a robot's extroversion can affect proxemics, a potential indicator of trust \cite{moujahid2023come}. Additionally, research has highlighted how a poorly designed robot personality can undermine trust \cite{kaniarasu2014effects}, \cite{giorgi2022friendly}. However, overall, the field remains fragmented \cite{mou2020systematic}. 

The connection between \textit{robot personality} and \textit{empathy} remains largely unexplored as well. Existing studies typically aim to elicit empathy toward robots solely through facial expressions, overlooking the roles of memory, personality, and prospective thinking \cite{park2022empathy}.

%Rationale behind this research
This work aims to investigate the unexplored link between \textit{Trust}, \textit{Empathy}, and \textit{Robot personality}. %with the final aim to exploit robotic personality to build companionship by improving trust and empathy. 
To this end, we employ a task- and platform-independent cognitive architecture built on the concept of robotic personality \cite{nardelli2023software}, \cite{nardelli2024personality}, \cite{nardelli2024personality2}, \cite{nardelli2025toward} (Section \ref{sec:arc}). First, we carefully designed and validated a user emotion-aware emotion generation component, as the ability to display emotion is a prerequisite for triggering empathy (Section \ref{sec:emgen}). Next, we conducted a user study involving 84 dyadic conversation sessions with the humanoid robot Navel (Section \ref{sec:sd}). Finally, a multimodal analysis of the results allowed us to validate the emotion generator and derive guidelines on leveraging robotic personality to foster companionship by enhancing trust and empathy (Section \ref{sec:res})

\section{Methods}
\vspace{-3mm}
\label{sec:arc}
 \begin{figure}[h!tb]
    \centering
    \includegraphics[width=0.5\textwidth]{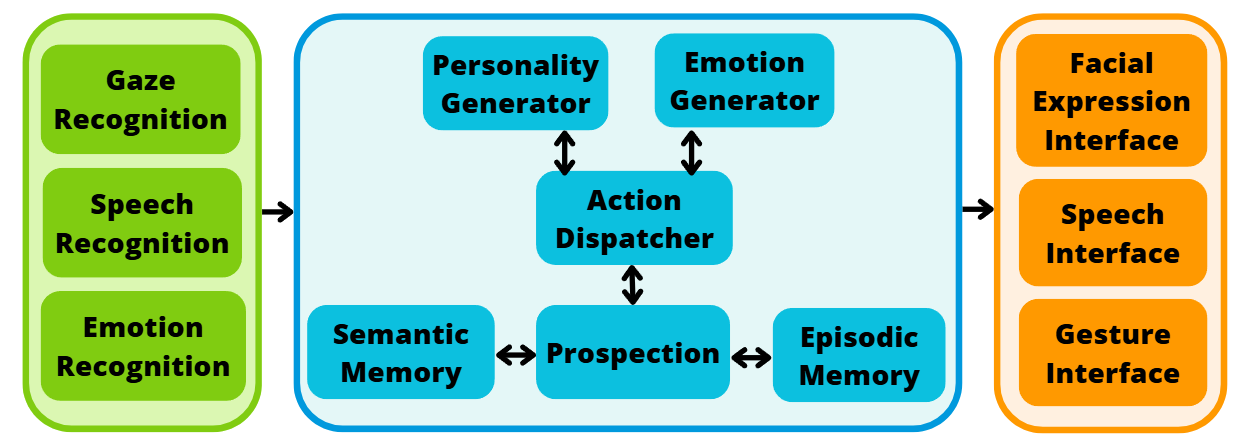}
    \caption{Personality-based cognitive architecture. Three different modules can be identified: perception (green), reasoning (blue), and actions (orange).}
    \vspace{-3mm}
    \label{fig:architecture}
\end{figure}

Figure \ref{fig:architecture} illustrates the cognitive architecture for robotic personality, previously detailed in our research \cite{nardelli2023software, nardelli2024personality, nardelli2024personality2, nardelli2025toward}. It has been implemented in various contexts, including the humanoid robot Pepper for a collaborative game \cite{nardelli2023software, nardelli2024personality}, a Digital Human in dyadic interactions \cite{nardelli2024personality2}, and the Kinova Jaco2 arm for human-robot collaboration \cite{nardelli2025toward}. For this study, we adapted the architecture for general-purpose conversation with the humanoid robot Navel (Figure \ref{fig:setup}), leveraging its task- and platform-independent design. This section provides an overview of its main components.

%The overall communication system is managed through a REST-API-based server. This cloud-based system (represented by all blue blocks in Figure \ref{fig:architecture}, while the green and orange blocks are intended for implementation on client applications running onboard robots or artificial agents) is compatible with various programming languages and can be easily adapted to different tasks and platforms, enhancing the framework's flexibility and usability.

\subsection{Perception}
\label{sec:percep}
The green perception blocks implement sentence recognition (via Microsoft Azure\footnote{\url{https://learn.microsoft.com/en-us/azure/ai-services/speech-service/}}) and facial emotion and gaze behavior recognition (using the Navel API). To handle emotions, we detect Ekman’s six basic emotions \cite{ekman1992there} along with Neutral. Facial emotions are identified based on the most probable expression within a 5-second sliding window. We employ a multimodal emotion-recognition approach, combining facial emotion detection with sentiment analysis of human speech using the GPT-4o language model\footnote{\url{https://openai.com/}}.%It is important to clarify that our objective is not to accurately understand human emotions but rather to develop a system capable of building an emotionally intelligent cognitive framework.

\subsection{Personality Definition}
\label{sec:per}
We model robotic personality as a vector in a three-dimensional space defined by Conscientiousness, Extroversion, and Agreeableness (CEA), allowing the generation of diverse personality profiles \cite{nardelli2023software}. 

$$
Personality= W_{c}C +  W_{e}E +  W_{a}A \eqno{(1)}
$$

where \(C\), \(E\), and \(A\) are the unit vectors of the three axes, and \(W_{c}\), \(W_{e}\), and \(W_{a}\) represent the degree to which each trait is expressed, ranging from \([-1 (low), +1 (high)]\) with \(0\) indicating neutrality. The model’s effectiveness has been validated across different platforms and tasks \cite{nardelli2023software, nardelli2024personality, nardelli2024personality2, nardelli2025toward}.

\subsection{Personality Generator}
\label{sec:personalitygen}

%yet robotic studies often rely on task-specific parameters, resulting in rigid and non-generalizable implementations. In contrast, human personality consists of universal traits that shape behavior across different contexts, even if not all traits manifest in every action.

To model how personality influences action execution, we integrated language generation techniques into the Personality Generator, leveraging the BERT attention-based architecture \cite{devlin2018bert}. The personality model takes a defined personality (Equation 1) and a general action as input, generating parameters that shape the action according to personality traits. These parameters include voice pitch, velocity, volume, language style, gaze behavior, gesture speed and amplitude, head movements, navigation speed, and proxemics.

Building on psychological insights, prior research \cite{nardelli2023software} identifies behavioral parameters influenced by the CEA model, constructs a dataset linking actions to CEA traits, and fine-tunes BERT accordingly. In this study, not all parameters are applicable, so we focus on gaze behavior (mutual or avoidant), gesture amplitude (high, low, or middle), voice volume, head movements, and language style \cite{nardelli2024personality2}. Specifically, Extroverts tend to be friendly, talkative, and enthusiastic,  while Introverts are reserved, quiet, and neutral. Agreeable personalities exhibit cooperativeness, friendliness, empathy, forgiveness, reliability, and politeness, whereas Disagreeable ones are competitive, aggressive, provocative, selfish, and rude. Conscientious individuals are scrupulous and precise, while Unconscientious ones tend to be thoughtless, distracted, lazy, and disorganized. For a detailed description of the full mapping between behavioral and robotic parameters we refer readers to our previous work \cite{nardelli2023software}.

\subsection{Prospection, Semantic Memory, and Episodic Memory}
\label{sec:mem}

Semantic Memory stores a personality-independent representation of the world through an ontology implementation (i.e., a set of propositions and predicates used to describe the world, which the system retrieves as needed) that allows us to capture its symbolic nature \cite{vernon2014artificial}.

 \begin{figure}[h!tb] 
    \centering
    \includegraphics[width=0.5\textwidth]{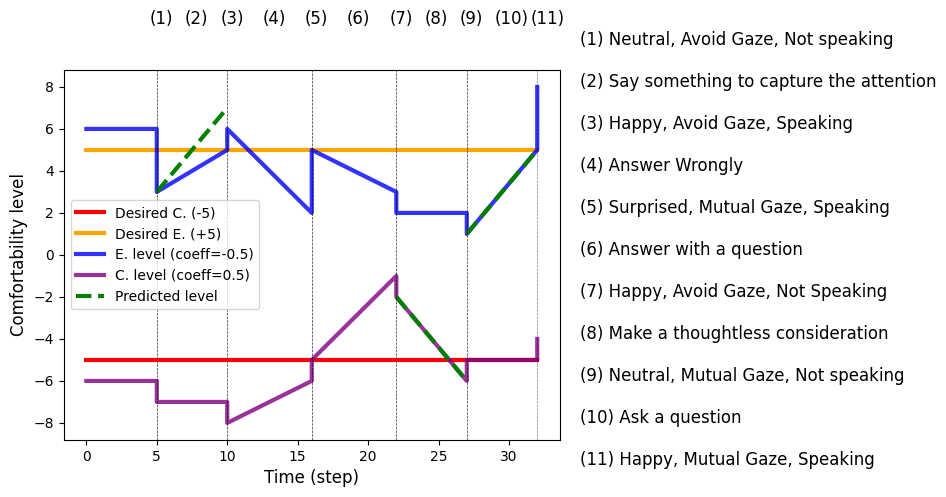}
    \caption{The figure illustrates personality-dependent functions (in purple and blue) and their responses to various actions, showing possible plans developed for an extrovert and unscrupulous agent, the selection of actions, their evaluation, and updates to the Episodic Memory. The contrast between the green dashed lines (indicating expected outcomes) and the solid lines (representing actual outcomes) highlights the gap between the anticipated reward and the obtained one.}
    \vspace{-5mm}
    \label{fig:prosp}
\end{figure}

Prospection, or internal simulation, enables individuals to plan future actions while satisfying hedonic needs. Due to its complexity, many cognitive architectures omit this component \cite{sandini2021cognitive}.
To address this, we propose a novel implementation \cite{nardelli2024personality}, \cite{nardelli2024personality2}, \cite{nardelli2025toward} that integrates Semantic Memory with a Fast-Forward (FF) planner \cite{hoffmann2001ff}, using an iterative planning strategy. Within the planning domain, for each CEA trait, we introduce numeric fluents to model Comfortability (Figure \ref{fig:prosp}, blue and purple lines). The variation in Comfortability depends on the agent's personality definition (Equation (1)) \cite{nardelli2024personality}, \cite{nardelli2025toward}. The planner ensures that the absolute Comfortability (Figure \ref{fig:prosp}, red and yellow lines) remains above a set threshold by introducing it as an action precondition.

Through this mechanism, the planner can simulate the variation of Comfortability, regulate it with a predictive homeostasis (allostasis) behavior \cite{vernon2014artificial}, and select the action path that best fulfills the hedonic experience of an agent with a specific personality defined within the CEA taxonomy. For instance, an extraverted agent in conversation may choose to answer with questions to sustain engagement or say something to capture attention if Comfortability drops below the threshold (Figure \ref{fig:prosp}, actions (2), (10)).

Episodic Memory further refines this process by storing past experiences and outcomes \cite{nardelli2024personality2}. It reinforces actions that elicit specific emotional responses; for example, a disagreeable agent may favor actions that provoke anger. The agent evaluates whether the actual outcome (e.g., user emotions, gaze) aligns with predictions, adjusting Comfortability accordingly (Figure \ref{fig:prosp}, actions (2), (8), (10)). Furthermore, Comfortability is directly influenced by environmental stimuli such as user emotions and gaze, reflecting personality sensitivity to external cues \cite{pease2015personality}, 
\cite{deyoung2020personality}. As it varies, the agent may take motivational goal actions without a direct practical objective, enhancing its perceived proactivity (Figure \ref{fig:prosp}, actions (1-2)).

\subsection{Emotion Generation}
\label{sec:emgen}
 
 \begin{figure*}[h!tb] 
    \centering
    \includegraphics[width=0.8\textwidth]{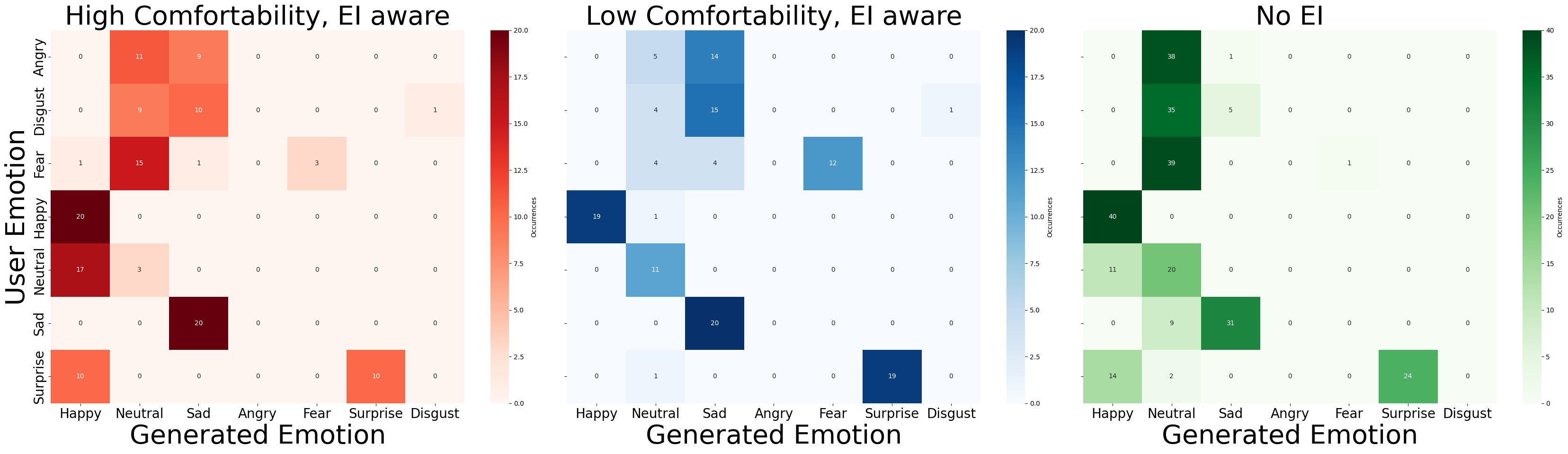}
    \caption{Evaluation of the Robot Emotion Generator, for an agreeable robot in response to different user's emotions. The three subplots distinguish the different robot's conditions (with and without emotional intelligence (EI), comfortable or uncomfortable).}
    \label{fig:seven_plots}
    \vspace{-3.9mm}
\end{figure*}

Emotional intelligence in humans, particularly emotional reactivity, strongly correlates with Big Five traits \cite{siegling2015incremental}, 
\cite{petrides2010relationships}, 
\cite{vernon2008phenotypic}. 
Studies \cite{han2012robotic}, \cite{masuyama2018personality} have explored this connection by developing a model-based emotion generator incorporating Big Five traits, perceptual stimuli, and associative memory. In contrast, this research proposes an emotion generator that utilizes the GPT-4o model to capture emotionally intelligent behaviors linked to each of the CEA traits. This approach addresses the task- and platform-independence of the architecture, as it can be easily integrated with new perceptual stimuli, diverse emotion classifications, and various cognitive processes.

We inform the GPT-4o model through the system prompt that its task is to serve as the emotion generator for an agent with a specific personality trait, describing the trait-related emotional intelligence in natural language \cite{vernon2008phenotypic}. The actual input includes the following fields: text, which contains what the user said; user emotion, containing the detected user emotion (Section \ref{sec:percep}); and comfortability, indicating whether the robot feels comfortable or uncomfortable (Section \ref{sec:mem}).

To validate our approach—assessing whether incorporating comfortability and the robot’s personality-related emotional intelligence description adds value—we compared emotion generation performance between the described prompt (EI aware condition) and a simpler prompt (NoEI condition) for an agreeable robot. The NoEI condition only included the user's utterance, the user’s emotion, and the robot's personality, without considering the emotional intelligence behaviors associated with the personality traits or the robot’s comfortability. We then created a set of inputs: for each of Ekman’s emotions \cite{ekman1992there}, we used GPT-4o to generate 40 sentences that could be expressed by a person feeling that specific emotion. We set the robot's comfortability level to comfortable for the first 20 sentences and to uncomfortable for the remaining 20.

Figure \ref{fig:seven_plots} shows the obtained results. We found that the NoEI condition predominantly generated neutrality. In contrast, the EI aware condition generated more accurate and nuanced emotions. For example, when the user is perceived as afraid or disgusted, the prompt of the EI aware condition generates sadness, better reflecting agreeableness and empathic traits (Figure \ref{fig:seven_plots}, Fear, Disgust user’s emotion). Moreover, the EI aware condition also takes into account the robot's internal state, generating more negative valence emotions in cases of discomfort (Figure \ref{fig:seven_plots}, Neutral user’s emotion).

In the practical integration of the generator into the cognitive architecture, since the system can manage three traits concurrently, we randomly select one specific trait, considering different weights associated with each personality pole. These weights depend on the robot's actual personality (as defined by the coefficient in Equation (1)) and the sensitivity of each pole to the perceived emotion \cite{canli2001fmri}, \cite{pease2015personality}, \cite{deyoung2020personality}. In this way, and in line with psychological literature \cite{deyoung2020personality}, emotions are more frequently generated based on the emotional intelligence behaviors associated with the traits of agreeableness and extraversion.

The robot's emotions are used as input in sentence generation (Section \ref{sec:exe}) and for the robot's facial expressions.

\subsection{Action Dispatcher} 
The Action Dispatcher regulates information flow between components by coordinating perception and action modules to trigger execution accordingly. Notably, it continuously extracts human emotions, even in the absence of direct interaction, and updates the Prospection module (Section \ref{sec:mem}). This ongoing process allows the agent to initiate actions when needed, ensuring a proactive rather than passive response, and maintaining an engaging and context-aware interaction.

\subsection{Execution Blocks}
\label{sec:exe}
The orange execution blocks, as shown in Figure \ref{fig:architecture}, control personality-influenced behaviors such as speech, gaze, gesture amplitude, head movements, and facial expressions. Sentence generation utilizes the GPT-4 model, with the system prompt instructing the model to generate sentences for the robot Navel based on a variety of factors, whose values are provided with each request. These factors include the user's emotion, the user's utterance, the robot's personality (Section \ref{sec:per}), the language style for sentence generation (Section \ref{sec:personalitygen}), the action generated by prospection (Section \ref{sec:mem}), and the robot's current emotion (Section \ref{sec:emgen}). An example of sentence generation for the robot is provided below:

\textbf{Input: \textit{Human sentence}: What did you say?, \textit{Human emotion}: Happy, \textit{Robot personality}: Disagreeable and Unscrupolous, \textit{Language style}: Thoughtless and Provocative, \textit{Action}: Make an affirmation, \textit{Robot emotion}: Disgust} 

\textbf{Output: \textit{robot sentence}: I said that the cats have taken over the space base, and I don’t really care about that.}

\section{Study design}
\label{sec:sd}
This section is fully dedicated to describe the design followed to conduct the user study. 

\subsection{Hypothesis}
\label{sec:h}
The focus of this research is to conduct a user study that addresses the following research questions:

\begin{itemize} 
\item \textit{Is the system capable of generating emotions accurately based on the robot's personality?} 
\item \textit{Is variation in synthetic personality perceivable across all CEA traits?} 
\item \textit{Does the robot's personality affect the perception of Experience, Empathy, Trust and the overall likability of the conversation?} 
\end{itemize}

Therefore, we formulated the following hypotheses:

\begin{itemize} \item \textit{Hypothesis 0}: The system is capable of generating personality-dependent emotions during real interactions with users. \item \textit{Hypothesis I}: Users perceive variations in personality across the three traits, even when they are exhibited concurrently. \item \textit{Hypothesis II}: Variations in agreeableness influence the perception of experience. \item \textit{Hypothesis III}: Variations in agreeableness and extraversion influence perceived empathy, while variations in conscientiousness, being less responsive to others' emotions, do not affect perceived empathy. \item \textit{Hypothesis IV}: Capability is primarily influenced by conscientiousness, while relational trust is influenced by agreeableness. \item \textit{Hypothesis V}: High levels of agreeableness and extraversion enhance both enjoyability and sociability. \end{itemize}

\subsection{Experimental set-up}
 \begin{figure}[h!tb]
    \centering
    \includegraphics[width=0.25\textwidth]{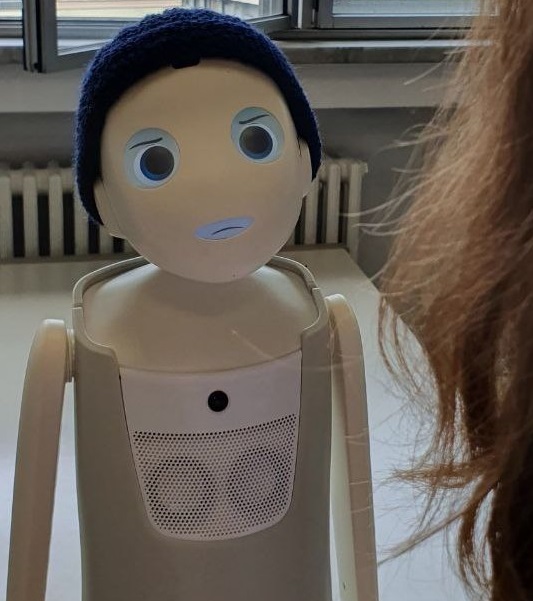}
    \caption{Experimental set-up}
    \label{fig:setup}
    \vspace{-3mm}
\end{figure}
The social robot Navel (Figure \ref{fig:setup}), developed by Navel Robotics\footnote{\url{https://navelrobotics.com/en/home-en-2/}}, was selected for the experiment due to its distinctive ability to display highly accurate facial expressions, thanks to the three-dimensional lenses above the OLED displays and the fine movements of the head that maintain mutual gaze. Therefore, it serves as an ideal platform for testing the emotional intelligence capabilities of our framework.

The framework has been designed to manage free-form conversations (Figure \ref{fig:architecture}), where the robot can autonomously guide the dialogue. The robot's personality is expressed through its actions, how it interprets others' emotions, its gaze behavior, decision-making processes, and the hedonic experiences linked to these interactions. Choosing a dyadic conversation as an open task allows participants to explore the empathetic cues of the agent without being confined to a specific conversational topic, thereby minimizing bias in our data \cite{inoue2022can}. Participants are instructed only not to speak while the robot is talking.

To determine the optimal number of participants required to achieve the desired effects outlined in Section \ref{sec:h}, as well as the appropriate number of trials per participant, we conducted a power analysis %\cite{erdfelder1996gpower}. 
We selected the Mann-Whitney U test \cite{nachar2008mann} a priori as the statistical method for analyzing the questionnaires, aiming to assess whether two samples belong to different distributions. We specified a two-tailed test with a normal parent distribution, setting the significance level at $\alpha=0.05$ and power at $1-\beta=0.95$. To calculate the effect size $d$ for personality perception and the variation in Agency, Experience, Enjoyability, and Sociability due to changes in personality along one of the three dimensions, we used the means and standard deviations from existing user studies \cite{mileounis2015creating}. In the worst-case scenario, we identified a sample size of 25 participants per group with an actual power of 0.957. To estimate the effect size $d$ for Trust due to variation in personality, we computed the effect size using data from prior research \cite{lim2022we}. Based on their results, we determined a sample size of 26 participants per group with an actual power of 0.950.

For empathy, we hypothesized a statistically significant result based on a 0.8 variation along the axis of interest (e.g., mean group 1 = 3, mean group 2 = 3.8), with a standard deviation of 0.75 for both groups. The results indicated that a sample size of 25 participants was needed to achieve an actual power of 0.950. We note that the proposed framework manages multiple personality traits simultaneously. For this reason, we decided to test only the extremes of the different personality traits, combining them into pairs, which resulted in 12 distinct personalities to test. Consequently, we opted for a fractional factorial partial within-subjects design (synthetic personality), where each participant engages in three sessions, experiencing three different combinations of personality traits. These combinations are shuffled in advance to balance the 12 personality traits and minimize order effects \cite{collins2009design}. By recruiting 28 participants, with each completing all three sessions, we ensured 28 data points per condition (e.g., extravert vs. introvert). This is the smallest number exceeding the sample size identified by the power analysis, sufficient to balance the 12 personality combinations while addressing the effects of testing multiple traits simultaneously.

We recruited indeed a total of 28 participants (27.39 ± 1.24 years old; 13 females, 15 males). Participants were recruited through online platforms and by word of mouth at the University of Genoa. They were informed that they were participating in an experiment involving verbal interaction with the robot Navel. Upon entering the laboratory, participants were required to sign an informed consent form that explained the experiment and outlined social permissions. They then completed a demographic questionnaire and were informed that they would engage in free conversation with Navel for three trials, each lasting 5 minutes. At the end of each trial, they were required to complete a questionnaire. Each participant participated in three sessions of free conversation with Navel, experiencing a different combination of traits in each session. Consequently, each of the 12 personalities was tested 7 times, with each personality pole being tested 28 times.

\vspace{-1mm}
\subsection{Measurement}
\subsubsection{Questionnaires}
To validate the presented hypotheses, participants completed a 5-point Likert scale questionnaire at the end of each interaction, assessing their impression of the robot's behavior during the session. The questionnaire included:
\begin{itemize}
    \item the Italian-validated version of the 10-item Big Five Inventory \cite{guido2015italian}, suitably adapted to the third person to assess perceived synthetic personality (\textit{Hypothesis I});
    \item the Agency and Experience questionnaire \cite{gray2007dimensions} to understand the perceived robot capability of feeling emotion, and the related effect of personality (\textit{Hypothesis II});
    \item the RoPE questionnaire, designed to assess perceptions of the robot's empathy, with two different scales: Empathic Understanding (EU) and Empathic Response (ER) %\cite{birmingham2022perceptions} 
    \cite{charrier2019rope} (\textit{Hypothesis III});
    \item the MDMT questionnaire \cite{ullman2018does}, which captures the multidimensionality of trust in a robot (\textit{Hypothesis IV});
    \item the Enjoyability and Sociability items of the UAUT questionnaire to investigate the enjoyment of interacting with the robot (\textit{Hypothesis V}).
\end{itemize} 

The questionnaires were administered online through the open-source SoSci Survey platform \footnote{\url{https://www.soscisurvey.de/de/index}}. The items of the questionnaire were randomly shuffled to prevent any order effects, and an attention-check item ("Answer 5 to this question to demonstrate your attention") was included to ensure that participants remained focused on the task.

\subsubsection{Behavioral data}
In order to validate the robot's capability to generate personality-dependent emotions, we recorded the robot's emotions and sentences during all the sessions (\textit{Hypothesis 0}). To further investigate how the robot's behavior influences the user's emotional state, we recorded the user's sentences and facial expressions. Additionally, we recorded user gaze behavior to determine whether it was mutual or averted, aiming to explore the relationship between trust and gaze behavior \cite{zhang2024gaze}.

\subsubsection{Open questions}
Participants answered the open-ended question, \textit{Which emotions can Navel feel?}. This provided a nuanced evaluation of the robot’s perceived ability to experience emotions and its connection to personality.

\section{Results}
\label{sec:res}
\subsection{Questionnaire analysis}
Before conducting the statistical analysis, we assessed the reliability of the employed scales by calculating Cronbach's alpha \cite{cronbach1951coefficient}. All scales achieved a coefficient greater than 0.7, confirming their reliability and internal consistency.

We analyzed all scales using the Mann-Whitney U test to determine whether variations between two distributions were statistically significant \cite{nachar2008mann}. In our analysis, the agent's opposite personality poles within the same trait served as independent variables, while the variable of interest was the dependent variable.

\subsubsection{Perceived Robot personality}
 \begin{figure*}[h!tb]
    \centering
    \includegraphics[width=0.8\textwidth]{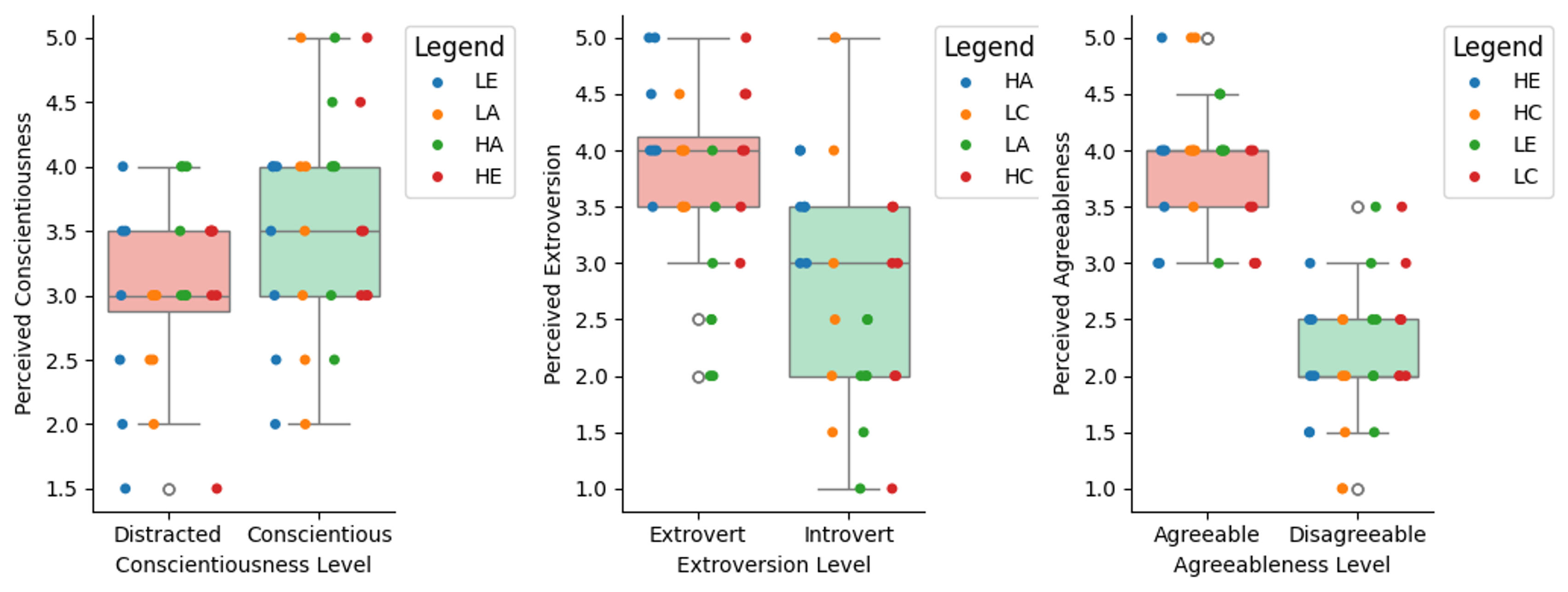}
    \caption{Perception of the different traits of the CEA taxonomy during the interaction with the robot Navel. E, A, C are the three personality traits. L and H mean respectively high value and low value of the dimension of interest. (e.g. LE means introversion and HE means extraversion). The hues indicate how the specific pole of interest is combined with the poles of the remaining two traits in the taxonomy.}
    \vspace{-3.9mm}
    \label{fig:boxplot}
\end{figure*}

We explored the robot's capability to convey personality by analyzing the Extroversion, Conscientiousness, and Agreeableness items of the BFI-10 questionnaire separately. The results validate \textit{Hypothesis I}, showing that participants distinguish, in a statistically significant way, between introvert and extravert robots ($p<0.001$), agreeable and disagreeable robots ($p<0.001$), and conscientious and unscrupulous robots ($p=0.027$). The box plot in Figure \ref{fig:boxplot} illustrates participants' perception of the three personality traits.

\subsubsection{Perceived Robot Agency and Experience} \label{sec:rex}

Regardless of personality, the robot is perceived as a cognitive agent capable of planning actions and recognizing users' emotions. Agreeableness is the only trait that significantly influences the robot’s perceived experience ($p<0.001$), confirming \textit{Hypothesis II}. The mean experience values (Extravert: $3.19 \pm 0.588$, Introvert: $3.12 \pm 0.857$, Agreeable: $3.47 \pm 0.458$, Disagreeable: $2.71 \pm 0.703$, Conscientious: $3.13 \pm 0.756$, Unscrupulous: $3.26 \pm 0.587$) suggest that the disagreeable robot is perceived as less capable of feeling both positive and negative emotions.

\subsubsection{Perceived Robot Empathy} \label{sec:re}

The results show that only Agreeableness affects both EU and ER ($p<0.001$ in both tests). This partially confirms \textit{Hypothesis III}, indicating that, among the three considered traits, Agreeableness influences perceived empathy.

\subsubsection{Perceived Robot Trust}
\label{sec:rt}
We analyzed the four dimensions of the MTMD: Capable, Ethical, Sincere, and Reliable \cite{ullman2018does}. The results show that Extraversion does not influence any trust dimension, while Conscientiousness affects only Capability ($p=0.043$). This suggests that participants perceive the unscrupulous agent as less capable, though this does not impact relational trust toward Navel. In contrast, Agreeableness influences all trust dimensions (Capable: $p=0.01$; Ethical, Sincere, and Reliable: $p<0.001$), with participants showing no trust in a disagreeable robot. This partially confirms \textit{Hypothesis IV}, as Agreeableness also affects trustworthiness. %In summary, Agreeableness influences both relational and capacity trust. This is likely because the prospection process amplifies the discomfort associated with a disagreeable robot, as it often appears reluctant to respond to the user. Conversely, a distracted robot may not inspire users’ capacity trust, but users are still willing to build a relationship with it.

\subsubsection{Perceived Robot Enjoyability and Sociability}
The UTAUT questionnaire results show that variations in Agreeableness significantly affect both Enjoyability and Sociability ($p<0.001$). While Extraversion influences Enjoyability, the effect does not reach statistical significance ($p=0.06$), preventing confirmation of \textit{Hypothesis V}. However, the comparable mean values for highly extraverted and highly agreeable personalities suggest that both traits contribute to increased enjoyment (Extravert: $3.84 \pm 0.694$, Introvert: $3.39 \pm 1.006$, Agreeable: $3.94 \pm 0.762$, Disagreeable: $2.98 \pm 0.989$, Conscientious: $3.43 \pm 1.012$, Unscrupulous: $3.58 \pm 0.965$).

\subsection{Behavioral Data}
\subsubsection{Robot Data Analysis}
The framework's ability to generate accurate emotions was validated by recording the occurrence of each emotion across experimental trials and computing the mean. A Mann-Whitney U test \cite{nachar2008mann} was performed for each emotion and personality trait, using the two opposite robotic personality poles as independent variables. Results indicate that extravert robots expressed more surprise and happiness than introverted ones ($p=0.02$, $p<0.001$), while introverted robots exhibited more neutral expressions ($p=0.008$). The unscrupulous robot showed higher levels of surprise, sadness, and disgust compared to the conscientious robot ($p<0.001$, $p=0.005$, $p=0.029$), which remained predominantly neutral. Agreeable robots were happier, sadder, and more fearful ($p<0.001$, $p<0.001$, $p=0.04$), whereas disagreeable robots displayed more anger and disgust ($p<0.001$, $p<0.001$).

These findings support the personality-dependent emotion generation system (\textit{Hypothesis 0}), demonstrating emotional intelligence linked to CEA traits \cite{deyoung2020personality}.

% which identified that extraversion is linked to experiencing high arousal, especially positive emotions. Highly agreeable agents tend to feel discomfort in conflict situations, while low-agreeable robots may seek out conflicts. Finally, conscientiousness is associated with a greater ability to control immediate impulses.

\subsubsection{Human Data Analysis}
The human data analysis involved recording facial expressions and gaze behavior (mutual vs. averted gaze) following the same procedure as the robot data analysis. Results showed that robot personality did not influence human gaze behavior or facial expressions; participants generally maintained a neutral expression and made eye contact with the robot.

Additionally, human speech was recorded and analyzed for emotion classification using GPT-4. The results indicated that the Agreeable robot elicited more happiness and surprise from users ($p<0.001$, $p=0.06$), while the Disagreeable robot prompted increased expressions of anger, sadness, and disgust ($p<0.001$, $p=0.02$, $p=0.002$). The Conscientious robot led to more happy sentences ($p=0.034$), whereas the Distracted robot increased expressions of surprise ($p=0.006$). These findings confirm the framework’s ability to elicit human emotions, with the empathic robot fostering positive emotions and the disagreeable robot triggering negative responses, validating the personality-dependent emotional responsiveness (\textit{Hypothesis III}). Furthermore, the fact that both highly conscientious and agreeable robots increased expressions of happiness suggests they may have been perceived as more trustworthy \cite{ullman2018does}, reinforcing \textit{Hypothesis IV}.

Finally, we analyzed the number of sentences and words spoken by users to investigate a possible link between self-disclosure and trust. While personality did not affect the total number of words spoken, users interacting with the introvert, disagreeable, or distracted robots tended to produce a higher number of sentences ($p<0.001$, $p=0.018$, $p=0.017$). However, given that all conversations lasted 5 minutes, this is likely due to these robots using shorter sentences, prompting users to speak more.

\subsection{Open question}

\begin{figure}[h!tb]
    \centering
    \includegraphics[width=0.5\textwidth]{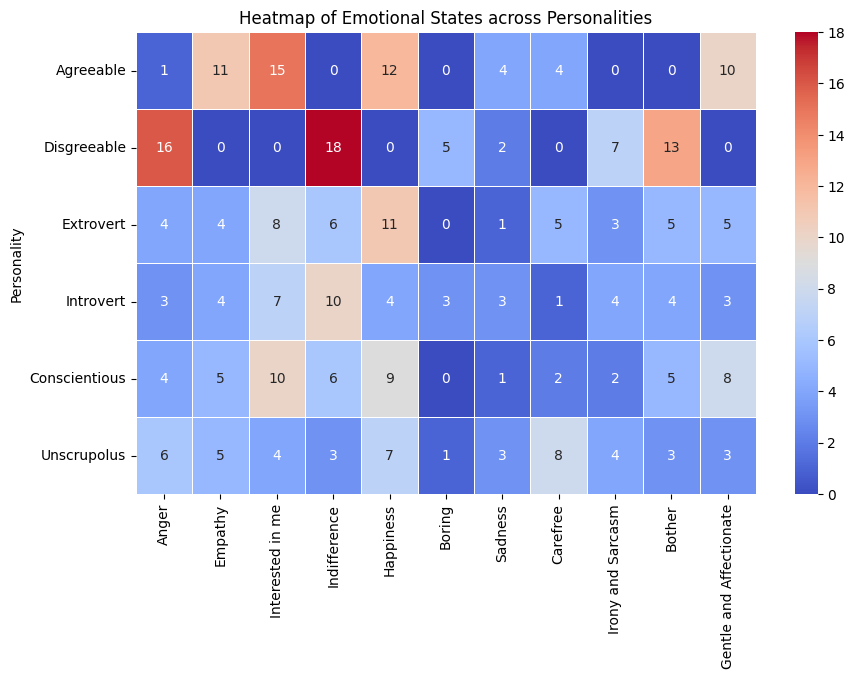}
    \caption{Heatmap showing, for each personality pole, the occurrencies of each emotional state}
    \vspace{-3.9mm}
    \label{fig:heatmap}
\end{figure}

To systematically identify the recurring themes in the responses to the open question \textit{Which emotions can Navel feel?}, and its link with the robotic personality, we followed the procedure of thematic analysis outlined by \cite{braun2012thematic}.
We identifyed the most frequent emotional states expressed by participants. Then, we counted the occurrences of each emotional category for each personality pole. To better visualize obtained results we organized them in a heatmap (Figure \ref{fig:heatmap}). The agreeable robot is the most capable of eliciting empathic emotional states, whereas the disagreeable robot is more likely to bother the users. The extravert robot is perceived as interested in the users and generally happy, while the introvert robot is perceived as less sociable. The conscientious robot is viewed as interested, happy, and gentle. Finally, the distracted robot is perceived as happy and carefree. This reflects participants' ability to accurately perceive the personality-influenced emotional-intelligent behavior.

\section{Conclusions}
The primary goal of the work is to establish guidelines for building companionship between humans and robots by introducing artificial personalities. We draw results on the previously unexamined connection between the traits of Conscientiousness, Extraversion, and Agreeableness and their impact on Empathy, Trust, Enjoyability, and Sociability.  To achieve this, we conducted a user study in which participants engaged in open-ended conversations with the empathic robot, Navel, which exhibited a combination of the aforementioned personality traits.

The results validate the framework’s ability to generate perceivable personality across these three dimensions, even when the traits are combined and highlight the necessity of combining multiple personality traits to design a successfull HRI. %The system demonstrates the capability of GPT-4o model of generating accurates personality-dependent emotions. 
Our findings highlight the importance of agreeableness in enhancing trust, empathy, and enjoyability. While conscientiousness seems to influence capacity trust, participants still expressed a desire to form relationships with a distracted robot. Finally, extraversion, appears to have an impact on the enjoyability of the interaction.

\bibliographystyle{ieeetr}
\bibliography{biblioPROPER}

\begin{thebibliography}{10}

\bibitem{ahmed2024human}
E.~Ahmed, O.~Buruk, and J.~Hamari, ``Human--robot companionship: Current trends and future agenda,'' {\em International Journal of Social Robotics}, pp.~1--52, 2024.

\bibitem{zinina2023learning}
A.~Zinina, A.~Kotov, N.~Arinkin, and L.~Zaidelman, ``Learning a foreign language vocabulary with a companion robot,'' {\em Cognitive Systems Research}, vol.~77, pp.~110--114, 2023.

\bibitem{lu2021effectiveness}
L.-C. Lu, S.-H. Lan, Y.-P. Hsieh, L.-Y. Lin, S.-J. Lan, and J.-C. Chen, ``Effectiveness of companion robot care for dementia: a systematic review and meta-analysis,'' {\em Innovation in aging}, vol.~5, no.~2, p.~igab013, 2021.

\bibitem{hoffman2016robotic}
G.~Hoffman, S.~Bauman, and K.~Vanunu, ``Robotic experience companionship in music listening and video watching,'' {\em Personal and Ubiquitous Computing}, vol.~20, no.~1, pp.~51--63, 2016.

\bibitem{esterwood2021systematic}
C.~Esterwood and L.~P. Robert, ``A systematic review of human and robot personality in health care human-robot interaction,'' {\em Frontiers in Robotics and AI}, vol.~8, p.~748246, 2021.

\bibitem{robert2020review}
L.~P. Robert~Jr, R.~Alahmad, C.~Esterwood, S.~Kim, S.~You, Q.~Zhang, {\em et~al.}, ``A review of personality in human--robot interactions,'' {\em Foundations and Trends{\textregistered} in Information Systems}, vol.~4, no.~2, pp.~107--212, 2020.

\bibitem{rossi2020role}
S.~Rossi, D.~Conti, F.~Garramone, G.~Santangelo, M.~Staffa, S.~Varrasi, and A.~Di~Nuovo, ``The role of personality factors and empathy in the acceptance and performance of a social robot for psychometric evaluations,'' {\em Robotics}, vol.~9, no.~2, p.~39, 2020.

\bibitem{galatolo2023personality}
A.~Galatolo, I.~Leite, and K.~Winkle, ``Personality-adapted language generation for social robots,'' in {\em 2023 32nd IEEE International Conference on Robot and Human Interactive Communication (RO-MAN)}, pp.~1800--1807, IEEE, 2023.

\bibitem{lim2022we}
M.~Y. Lim, J.~D.~A. Lopes, D.~A. Robb, B.~W. Wilson, M.~Moujahid, E.~De~Pellegrin, and H.~Hastie, ``We are all individuals: The role of robot personality and human traits in trustworthy interaction,'' in {\em 2022 31st IEEE International Conference on Robot and Human Interactive Communication (RO-MAN)}, pp.~538--545, IEEE, 2022.

\bibitem{robb2023explanation}
D.~A. Robb, X.~Liu, and H.~Hastie, ``Explanation styles for trustworthy autonomous systems,'' in {\em 22nd International Conference on Autonomous Agents and Multiagent Systems 2023}, pp.~2298--2300, Association for Computing Machinery, 2023.

\bibitem{moujahid2023come}
M.~Moujahid, D.~A. Robb, C.~Dondrup, and H.~Hastie, ``Come closer: The effects of robot personality on human proxemics behaviours,'' in {\em 2023 32nd IEEE International Conference on Robot and Human Interactive Communication (RO-MAN)}, pp.~2610--2616, IEEE, 2023.

\bibitem{kaniarasu2014effects}
P.~Kaniarasu and A.~M. Steinfeld, ``Effects of blame on trust in human robot interaction,'' in {\em The 23rd IEEE international symposium on robot and human interactive communication}, pp.~850--855, IEEE, 2014.

\bibitem{giorgi2022friendly}
I.~Giorgi, F.~A. Tirotto, O.~Hagen, F.~Aider, M.~Gianni, M.~Palomino, and G.~L. Masala, ``Friendly but faulty: A pilot study on the perceived trust of older adults in a social robot,'' {\em IEEE Access}, vol.~10, pp.~92084--92096, 2022.

\bibitem{mou2020systematic}
Y.~Mou, C.~Shi, T.~Shen, and K.~Xu, ``A systematic review of the personality of robot: Mapping its conceptualization, operationalization, contextualization and effects,'' {\em International Journal of Human--Computer Interaction}, vol.~36, no.~6, pp.~591--605, 2020.

\bibitem{park2022empathy}
S.~Park and M.~Whang, ``Empathy in human--robot interaction: Designing for social robots,'' {\em International journal of environmental research and public health}, vol.~19, no.~3, p.~1889, 2022.

\bibitem{nardelli2023software}
A.~Nardelli, C.~Recchiuto, and A.~Sgorbissa, ``A software framework to encode the psychological dimensions of an artificial agent.,'' in {\em 2023 32nd IEEE International Conference on Robot and Human Interactive Communication (RO-MAN)}, pp.~1711--1718, IEEE, 2023.

\bibitem{nardelli2024personality}
A.~Nardelli, G.~Maccagni, F.~Minutoli, A.~Sgorbissa, and C.~T. Recchiuto, ``Personality-and memory-based framework for emotionally intelligent agents,'' in {\em 2024 33rd IEEE International Conference on Robot and Human Interactive Communication (ROMAN)}, pp.~769--776, IEEE, 2024.

\bibitem{nardelli2024personality2}
A.~Nardelli, A.~Sgorbissa, and C.~T. Recchiuto, ``Personality-and memory-based software framework for human-robot interaction,'' in {\em 2024 IEEE International Conference on Robotics and Automation (ICRA)}, pp.~17388--17394, IEEE, 2024.

\bibitem{nardelli2025toward}
A.~Nardelli, L.~Landolfi, D.~Pasquali, A.~Sgorbissa, F.~Rea, and C.~Recchiuto, ``Toward a universal concept of artificial personality: Implementing robotic personality in a kinova arm,'' {\em arXiv preprint arXiv:2501.06867}, 2025.

\bibitem{ekman1992there}
P.~Ekman, ``Are there basic emotions?,'' 1992.

\bibitem{devlin2018bert}
J.~Devlin, M.-W. Chang, K.~Lee, and K.~Toutanova, ``Bert: Pre-training of deep bidirectional transformers for language understanding,'' {\em arXiv preprint arXiv:1810.04805}, 2018.

\bibitem{vernon2014artificial}
D.~Vernon, {\em Artificial cognitive systems: A primer}.
\newblock MIT Press, 2014.

\bibitem{sandini2021cognitive}
G.~Sandini, A.~Sciutti, and D.~Vernon, ``Cognitive robotics,'' {\em Encyclopedia of Robotics; Springer: Berlin/Heidelberg, Germany}, 2021.

\bibitem{hoffmann2001ff}
J.~Hoffmann and B.~Nebel, ``The ff planning system: Fast plan generation through heuristic search,'' {\em Journal of Artificial Intelligence Research}, vol.~14, pp.~253--302, 2001.

\bibitem{pease2015personality}
C.~R. Pease and G.~J. Lewis, ``Personality links to anger: Evidence for trait interaction and differentiation across expression style,'' {\em Personality and Individual Differences}, vol.~74, pp.~159--164, 2015.

\bibitem{deyoung2020personality}
C.~G. DeYoung and J.~R. Gray, ``Personality neuroscience: Explaining individual differences in affect, behaviour and cognition.,'' pp.~323--346, 2020.

\bibitem{siegling2015incremental}
A.~Siegling, A.~K. Vesely, K.~Petrides, and D.~H. Saklofske, ``Incremental validity of the trait emotional intelligence questionnaire--short form (teique--sf),'' {\em Journal of personality assessment}, vol.~97, no.~5, pp.~525--535, 2015.

\bibitem{petrides2010relationships}
K.~V. Petrides, P.~A. Vernon, J.~A. Schermer, L.~Ligthart, D.~I. Boomsma, and L.~Veselka, ``Relationships between trait emotional intelligence and the big five in the netherlands,'' {\em Personality and Individual differences}, vol.~48, no.~8, pp.~906--910, 2010.

\bibitem{vernon2008phenotypic}
P.~A. Vernon, V.~C. Villani, J.~A. Schermer, and K.~Petrides, ``Phenotypic and genetic associations between the big five and trait emotional intelligence,'' {\em Twin Research and Human Genetics}, vol.~11, no.~5, pp.~524--530, 2008.

\bibitem{han2012robotic}
M.-J. Han, C.-H. Lin, and K.-T. Song, ``Robotic emotional expression generation based on mood transition and personality model,'' {\em IEEE transactions on cybernetics}, vol.~43, no.~4, pp.~1290--1303, 2012.

\bibitem{masuyama2018personality}
N.~Masuyama, C.~K. Loo, and M.~Seera, ``Personality affected robotic emotional model with associative memory for human-robot interaction,'' {\em Neurocomputing}, vol.~272, pp.~213--225, 2018.

\bibitem{canli2001fmri}
T.~Canli, Z.~Zhao, J.~E. Desmond, E.~Kang, J.~Gross, and J.~D. Gabrieli, ``An fmri study of personality influences on brain reactivity to emotional stimuli.,'' {\em Behavioral neuroscience}, vol.~115, no.~1, p.~33, 2001.

\bibitem{inoue2022can}
K.~Inoue, D.~Lala, and T.~Kawahara, ``Can a robot laugh with you?: Shared laughter generation for empathetic spoken dialogue,'' {\em Frontiers in Robotics and AI}, vol.~9, p.~933261, 2022.

\bibitem{nachar2008mann}
N.~Nachar {\em et~al.}, ``The mann-whitney u: A test for assessing whether two independent samples come from the same distribution,'' {\em Tutorials in quantitative Methods for Psychology}, vol.~4, no.~1, pp.~13--20, 2008.

\bibitem{mileounis2015creating}
A.~Mileounis, R.~H. Cuijpers, and E.~I. Barakova, ``Creating robots with personality: The effect of personality on social intelligence,'' in {\em Artificial Computation in Biology and Medicine: International Work-Conference on the Interplay Between Natural and Artificial Computation, IWINAC 2015, Elche, Spain, June 1-5, 2015, Proceedings, Part I 6}, pp.~119--132, Springer, 2015.

\bibitem{collins2009design}
L.~M. Collins, J.~J. Dziak, and R.~Li, ``Design of experiments with multiple independent variables: a resource management perspective on complete and reduced factorial designs.,'' {\em Psychological methods}, vol.~14, no.~3, p.~202, 2009.

\bibitem{guido2015italian}
G.~Guido, A.~M. Peluso, M.~Capestro, and M.~Miglietta, ``An italian version of the 10-item big five inventory: An application to hedonic and utilitarian shopping values,'' {\em Personality and Individual Differences}, vol.~76, pp.~135--140, 2015.

\bibitem{gray2007dimensions}
H.~M. Gray, K.~Gray, and D.~M. Wegner, ``Dimensions of mind perception,'' {\em science}, vol.~315, no.~5812, pp.~619--619, 2007.

\bibitem{charrier2019rope}
L.~Charrier, A.~Rieger, A.~Galdeano, A.~Cordier, M.~Lefort, and S.~Hassas, ``The rope scale: a measure of how empathic a robot is perceived,'' in {\em 2019 14th ACM/IEEE International Conference on Human-Robot Interaction (HRI)}, pp.~656--657, IEEE, 2019.

\bibitem{ullman2018does}
D.~Ullman and B.~F. Malle, ``What does it mean to trust a robot? steps toward a multidimensional measure of trust,'' in {\em Companion of the 2018 acm/ieee international conference on human-robot interaction}, pp.~263--264, 2018.

\bibitem{zhang2024gaze}
Y.~Zhang, A.~Yadav, S.~K. Hopko, and R.~K. Mehta, ``In gaze we trust: Comparing eye tracking, self-report, and physiological indicators of dynamic trust during hri,'' in {\em Companion of the 2024 ACM/IEEE International Conference on Human-Robot Interaction}, pp.~1188--1193, 2024.

\bibitem{cronbach1951coefficient}
L.~J. Cronbach, ``Coefficient alpha and the internal structure of tests,'' {\em psychometrika}, vol.~16, no.~3, pp.~297--334, 1951.

\bibitem{braun2012thematic}
V.~Braun and V.~Clarke, {\em Thematic analysis.}
\newblock American Psychological Association, 2012.

\end{thebibliography}

\end{document}